\documentclass[conference]{IEEEtran}
\usepackage{cite}
\usepackage{amsmath}
\usepackage{amssymb}
\usepackage{amsfonts}
\usepackage{algorithmic}
\usepackage{graphicx}
\usepackage{textcomp}
\usepackage[dvipsnames]{xcolor}

\usepackage{hyperref}
\hypersetup{pagebackref=false,breaklinks=true,colorlinks,bookmarks=false,urlcolor=blue,linkcolor=blue,citecolor=blue,pdftitle={An Adaptive Solver for Systems of Linear Equations},pdfauthor={Conrad Sanderson, Ryan Curtin}}

\usepackage[latin1]{inputenc}
\usepackage[labelfont=bf,textfont=normalsize]{caption}
\usepackage{enumerate}
\usepackage{mathtools}
\usepackage{xspace}
\usepackage{multirow}
\usepackage{booktabs}
\usepackage{algorithmic}
\usepackage{algorithm}
\usepackage{xfrac}
\usepackage{fancyvrb}  
\usepackage{balance}
\usepackage{adjustbox}

\usepackage[frozencache]{minted}  

\usepackage[british]{babel}
\usepackage{inconsolata}  

\graphicspath{{./}{./figures/}}

\def\Vec#1{{\boldsymbol{#1}}}
\def\Mat#1{{\boldsymbol{#1}}}

\def\ie{i.e.\xspace}

\begin{document}

\title{\mbox{An Adaptive Solver for Systems of Linear Equations}}

\author
  {
  Conrad Sanderson~{$^{\dagger\ddagger}$} and Ryan Curtin~{$^{\diamond}$}\\
  ~\\
  {\normalsize{$^\dagger$}  Data61~/~CSIRO, Australia}\\
  {\normalsize{$^\ddagger$} Griffith University, Australia}\\
  {\normalsize{$^\diamond$} RelationalAI, USA}
  }

\maketitle

\begin{abstract}

Computational implementations for solving systems of linear equations
often rely on a one-size-fits-all approach based on LU decomposition of dense matrices 
stored in column-major format.
Such solvers are typically implemented with the aid of the {xGESV} set of functions
available in the low-level LAPACK software,
with the aim of reducing development time by taking advantage of well-tested routines.
However, this straightforward approach does not take into account various matrix
properties which can be exploited to reduce the computational effort
and/or to increase numerical stability.
Furthermore, direct use of LAPACK functions can be error-prone for non-expert users
and results in source code that has little resemblance to originating mathematical expressions.
We describe an adaptive solver that we have implemented inside recent versions
of the high-level Armadillo C++ library for linear algebra.
The solver automatically detects several common properties of a given system
(banded, triangular, symmetric positive definite),
followed by solving the system via mapping to a set of suitable LAPACK functions best matched to each property.
The solver also detects poorly conditioned systems
and automatically seeks a solution via singular value decomposition as a fallback.
We show that the adaptive solver leads to notable speedups,
while also freeing the user from using direct calls to cumbersome LAPACK functions.

\end{abstract}

\vspace{0.5ex}


\section{Introduction}

Solving systems of linear equations is a fundamental computational task in numerous fields~\cite{Demmel_1997,Golub_2013},
including signal processing and machine learning.
The general form for a system of linear equations is expressed as:
%
\begin{equation}
\begin{array}{ccccccccccccccccc}
a_{11} x_1 & \mbox{+} & a_{12} x_2 & \mbox{+} & {~\cdots~} & \mbox{+} & a_{1n} x_n & \mbox{=} & b_1      \\
a_{21} x_1 & \mbox{+} & a_{22} x_2 & \mbox{+} & {~\cdots~} & \mbox{+} & a_{2n} x_n & \mbox{=} & b_2      \\
{\vdots}   & \mbox{~} & {\vdots}   & \mbox{~} & {\vdots}   & \mbox{~} & {\vdots}   & \mbox{~} & {\vdots} \\
a_{m1} x_1 & \mbox{+} & a_{m2} x_2 & \mbox{+} & {~\cdots~} & \mbox{+} & a_{mn} x_n & \mbox{=} & b_m
\end{array}
\label{eqn:expanded_form}
\end{equation}

\noindent
which can be compactly represented in matrix form as:
\begin{equation}
  \Mat{A} \Vec{x} = \Vec{b}
  \label{eqn:general_form}
\end{equation}

\noindent
where
matrix $\Mat{A}$ with size $m \times n$ contains the coefficients of the system,
vector $\Vec{x}$ with size $n \times 1$ represents the unknown variables to be found,
and
vector $\Vec{b}$ with size $m \times 1$ contains known constants.

For the common case of square-sized $\Mat{A}$,
a naive approach to find $\Vec{x}$ is via matrix inverse,
\ie~{$\Vec{x} = \Mat{A}^{-1} \Vec{b}$}.
However, in practice computing the inverse is actually not necessary
and can result in numerical instability, leading to inaccurate solutions~\cite{Higham_2002}.
An effective and general approach 
is to solve the system with the help of lower-upper (LU) decomposition~\mbox{\cite{Demmel_1997,Golub_2013}}.
This can be typically implemented through the {\tt xGESV} set of functions%
\footnote
  {
  For each LAPACK function, we substitute the first letter of the function with `{\tt x}' to indicate a set of functions
  which differ only in the applicable element type (eg.~single- or double-precision element type).
  For example, the set \{{\tt SGESV}, {\tt DGESV}, {\tt CGESV}, {\tt ZGESV}\}
  is represented as {\tt xGESV}.
  }
available in the ubiquitous and de-facto industry standard LAPACK software~\cite{anderson1999lapack}.
Several optimised versions of LAPACK are available,
such as MKL~\cite{IntelMKL} for the {\tt x86-64} architecture
used by the widely employed Intel and AMD processors.

The task of converting an arbitrary linear algebra expression
into an efficient sequence of optimally matched LAPACK function calls,
either manually or automatically,
is known as the {\it linear algebra mapping problem}~\cite{Psarras_2019}.
When the conversion is done manually,
the process can be laborious and error-prone;
it typically requires thorough understanding of several areas:
high-performance computing, numerical linear algebra, and the intricacies of LAPACK.
Furthermore, source code that directly uses LAPACK functions has several downsides: it
{\bf (i)}~has little resemblance to the originating \mbox{mathematical} expression,
{\bf (ii)}~requires manual memory management,
{\bf (iii)}~requires keeping track of many extra variables.
In turn, these downsides reduce the readability of the source code by \mbox{non-expert} users,
while increasing the maintenance burden and risk of bugs\mbox{~\cite{Sneed_2004,Malhotra_2016}}.

To address the above issues and hence to increase productivity,
linear algebra expressions are often implemented with the aid of high-level frameworks~\cite{Viviani_2018},
such as Matlab, Octave~\cite{Linge_MatlabOctave_2016} and Armadillo~\cite{Armadillo_JOSS_2016}.
While these frameworks facilitate simplified user code that focuses on high-level algorithm logic,
the automatic mapping of given mathematical expressions into LAPACK functions can be suboptimal~\cite{Barthels_2020,Berenyi_2018}.

Higher level frameworks default to storing dense matrices in straightforward column-major format
(where all the elements in each column are stored consecutively in memory),
without taking into account any special structure or properties of the matrix.
This choice is made by framework maintainers to reduce the internal code complexity of the framework,
and to avoid burdening users with the choice of storage types;
users may not have the inclination nor expertise to understand the advantages and disadvantages of various storage formats.

Within the above context, 
the dense matrix LU-based solver is an effective and often-used one-size-fits-all approach
for solving systems of linear equations.
However, this straightforward approach does not take into account various matrix properties which can be exploited
to reduce the computational effort and/or to increase numerical stability.
For example, when matrix $\Mat{A}$ is symmetric positive definite,
it is more computationally efficient to use Cholesky decomposition
instead of LU decomposition~\cite{Golub_2013}.

In this paper we describe an adaptive solver that we have devised and implemented
inside recent versions of the open-source Armadillo C++ linear algebra library.  
The library provides a high-level {\it domain specific language}~\cite{Mernik_2005,Scherr_2015} embedded within the host C++ language,
allowing mathematical operations with matrices to be expressed in a concise and easy-to-read manner similar to Matlab/Octave.
This facilitates prototyping directly in C++ and aids the conversion of research code into production environments.

The adaptive solver is able to automatically detect several common properties of matrices,
while being robust to inherent limitations in the precision of floating point representations~\cite{Goldberg_1991,Muller_2018},
and then map them to a large set of suitable LAPACK functions.
We show that this leads to considerable speedups,
thereby freeing the user from worrying about storage formats
and using cumbersome manual calls to LAPACK functions
in order to obtain good computational efficiency.

We continue the paper as follows.
Section~\ref{sec:mapping} describes several common matrix properties which can be exploited,
summarises the algorithms used for detecting such properties,
and lists the suitable sets of LAPACK functions tailored for each property.
Section~\ref{sec:experiments} provides an empirical evaluation
showing the speedups attained by the adaptive solver.
Section~\ref{sec:runtime} provides a brief discussion on computational complexity and runtime considerations.
The salient points and avenues for further exploitation are summarised in Section~\ref{sec:conclusion}.

\section{Automatic Detection and Mapping}
\label{sec:mapping}

\begin{figure*}[t!]
\centering
\includegraphics[width=1\textwidth,height=0.26\textwidth]{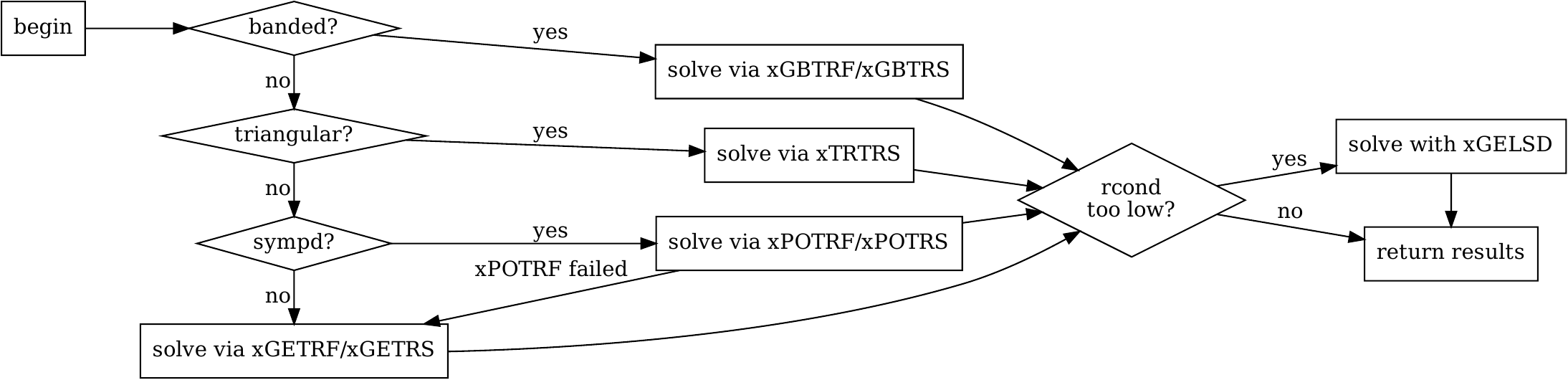}
\caption
  {
  Flowchart for the adaptive solver.
  Depending on properties of the input matrix $\Mat{A}$,
  more efficient LAPACK solvers are used when possible.
  In all cases, if the reciprocal condition number (rcond) of $\Mat{A}$ is too low,
  an approximate SVD-based solver is used as a fallback.
  }
\label{fig:flowchart}
\vspace{1ex}
\end{figure*}

The adaptive solver detects special matrix structures in the following order:
{\bf (i)}~banded, 
{\bf (ii)}~upper or lower triangular,
{\bf (iii)}~symmetric positive definite (sympd).
Examples of the structures are shown in Fig.~\ref{fig:matrix_structures}.
A specialised solver is employed as soon as one of the structures is detected.
The detection algorithms and associated LAPACK functions
for solving the systems are described in Sections~\ref{sec:banded_matrices} through to~\ref{sec:sympd_matrices}.
If no special structure is detected,
an extended form of an LU-based solver is used,
described in Section~\ref{sec:gen_matrices}.
A~flowchart summarising the adaptive solver is given in Fig.~\ref{fig:flowchart}.

\begin{figure}[!b]
\centering
\begin{minipage}{0.32\columnwidth}
\centering
\begin{equation*}
\footnotesize
\begin{bmatrix}
1 & 9 & 0 & 0 & 0 \\
6 & 2 & 8 & 0 & 0 \\
0 & 7 & 3 & 7 & 0 \\
0 & 0 & 8 & 4 & 6 \\
0 & 0 & 0 & 9 & 5
\end{bmatrix}
\end{equation*}
(a)
\end{minipage}
\hfill
\begin{minipage}{0.32\columnwidth}
\centering
\begin{equation*}
\footnotesize
\begin{bmatrix}
1 & 0 & 0 & 0 & 0 \\
2 & 6 & 0 & 0 & 0 \\
3 & 7 & 1 & 0 & 0 \\
4 & 8 & 2 & 4 & 0 \\
5 & 9 & 3 & 5 & 6
\end{bmatrix}
\end{equation*}
(b)
\end{minipage}
\hfill
\begin{minipage}{0.32\columnwidth}
\centering
\begin{equation*}
\footnotesize
\begin{bmatrix}
9 & 1 & 2 & 3 & 4 \\
1 & 8 & 1 & 2 & 3 \\
2 & 1 & 7 & 1 & 2 \\
3 & 2 & 1 & 6 & 1 \\
4 & 3 & 2 & 1 & 5
\end{bmatrix}
\end{equation*}
(c)
\end{minipage}
\caption
  {
  Examples of matrix structures:
  (a)~banded,
  (b)~lower triangular,
  (c)~symmetric positive definite.
  }
\label{fig:matrix_structures}
\end{figure}

Each of the solvers estimates the reciprocal condition number~\cite{Belsley_1980} of the given system,
which is used to determine the quality of the solution.
If any of the solvers fail to find a solution
(including the solver for general matrices),
or if the system is determined to be poorly conditioned, 
an approximate solution is attempted using a solver based on singular value decomposition (SVD),
described in Section~\ref{sec:svd_solver}.

The code for the detection and mapping is implemented as part of the {\small\tt solve()} function in Armadillo.
As of Armadillo version 9.900,
the {\small\tt solve()} function is comprised of about 3000 lines of code,
not counting LAPACK code.
Example usage of the {\small\tt solve()} function is shown in Fig.~\ref{fig:example_usage}.

\begin{figure}[!b]
\vspace{-2ex}
\centering
\hrule
\vspace{1ex}
\begin{adjustbox}{minipage=\columnwidth,scale={1.00}{1.00}}   
\begin{minted}[fontsize=\footnotesize,escapeinside=||]{c++}
#include <armadillo>

using namespace arma;

int main()
  {
  // generate matrix with random values in [-0.5,+0.5] interval
  mat R = |\textcolor{blue}{randu}|(100, 100) - 0.5;
  
  // generate symmetric matrix A via A = R'R
  mat A = R.t() * R;
  
  // ensure values on the the main diagonal are dominant
  A.diag() += 1.0;
  
  // generate random column vector 
  vec |\textcolor{black}{B}|(100, fill::randu);
  
  // solve for X in the random sympd system AX = B
  vec X = |\textcolor{blue}{solve}|(A,B);
  
  X.print("solution:");
  
  return 0;
  }
\end{minted}
\end{adjustbox}
\vspace{1ex}
\hrule
\vspace{1ex}
\caption
  {
  An example C++ program for solving a random sympd system.
  The {\small\tt solve()} function as well as the {\small\tt mat} and {\small\tt vec} classes
  (for matrices and vectors)
  are provided by the Armadillo library~\cite{Armadillo_JOSS_2016}.
  Several optional arguments have been omitted for brevity.
  The {\small\tt solve()} function is internally comprised of about 3000 lines of code,
  not counting LAPACK code.
  Documentation for all available classes and functions can be viewed at \href{http://arma.sourceforge.net/docs.html}{\small\tt http://arma.sourceforge.net/docs.html}.
  }
\label{fig:example_usage}
\end{figure}

\subsection{Banded Matrices}
\label{sec:banded_matrices}

Banded matrices contain elements on the main diagonal and typically on several more diagonals, above and/or below the main diagonal.
All other elements are zero.
As such, computational effort can be considerably reduced by exploiting the sparseness of the matrix.
Fig.~\ref{fig:matrix_structures}(a) shows an example banded matrix.

To efficiently detect a banded matrix structure within a column-major dense matrix,
each column is examined rather than each possible diagonal.
To determine the number of super-diagonals (diagonals above the main diagonal),
the elements in each column are traversed from the top of the column to the location of the main diagonal within that column,
followed by noting the difference in the locations of the first non-zero element and the main diagonal.
The maximum of all noted differences is taken as the number of super-diagonals.
To determine the number of sub-diagonals (diagonals below the main diagonal),
the elements in each column are traversed from the location of the main diagonal within that column to the bottom of the column,
followed by noting the difference in the locations of the main diagonal and the last non-zero element.
The maximum of all noted differences is taken as the number of sub-diagonals.

As soon as the number of detected diagonals indicates that more than 25\% of the elements within the matrix are non-zero,
all further processing is stopped and the matrix is deemed to be non-banded.
The threshold of 25\% was determined empirically as a good trade-off
between the lower computational requirements of solving a banded system,
and the amount of extra processing required for both the detection and further wrangling of data in the banded structure.

If a banded matrix structure is detected, data from the diagonals must be first converted into a relatively compact storage format,
where each diagonal is stored as a vector in a separate dense matrix~\cite{anderson1999lapack}.
Data in the compact storage format is then used by the LAPACK function {\tt xGBTRF} to compute 
the LU decomposition of the banded matrix,
followed by the {\tt xGBTRS} function to solve the system based on the LU decomposition,
and finally the {\tt xGBCON} function to estimate the reciprocal condition number from the LU decomposition.

Diagonal matrices are treated as banded matrices,
where the number of diagonals above and below the main diagonal is zero.
While it is certainly possible to have specialised handling for diagonal matrices,
in practice such matrices are seldom encountered when solving systems of linear equations.

\subsection{Triangular Matrices}
\label{sec:tri_matrices}

For triangular matrices, the LU decomposition can be avoided
and the solution can be obtained via either back-substitution (for upper triangular matrices)
or forward-substitution (for lower triangular matrices)~\cite{Golub_2013}.
As such, considerable reductions in computational effort can be attained.
An example of a lower triangular matrix is shown in Fig.~\ref{fig:matrix_structures}(b).

The process of forward-substitution can be summarised as first finding the value for {\small $x_{1}$} in Eqn.~(\ref{eqn:expanded_form}),
where $a_{12}$ through to $a_{1n}$ are known to be zero,
resulting in $x_{1} = b_{1} / a_{11}$.
The value for $x_{1}$ is then used to find $x_{2}$,
where $a_{23}$ through to $a_{2n}$ are zero.
This iterative process can be compactly expressed as:

\begin{equation}
x_{i} = \frac{b_{i} - \sum_{j=1}^{i-1} a_{ij} \cdot x_{j}}{a_{ii}}
\end{equation}%
The process of back-substitution follows a similar manner,
starting from the bottom of the matrix instead of the top
(\ie $x_{n}$ is solved first instead of $x_{1}$).

The detection of triangular matrices is done in a straightforward manner.
If all elements above the main diagonal are zero, a lower triangular matrix is present.
Conversely, if all elements below the main diagonal are zero, an upper triangular matrix is present.
An attempt to solve the system is made by using the {\tt xTRTRS} function,
and the reciprocal condition number is computed via the {\tt xTRCON} function.

\subsection{Symmetric Positive Definite Matrices}
\label{sec:sympd_matrices}

A matrix $\Mat{M}$ is considered to be symmetric positive definite (sympd)
if $\Vec{v}^{\top} \Mat{M} \Vec{v} > 0$ for every non-zero column vector $\Vec{v}$,
and $\Mat{M}$ is symmetric (\ie. its lower triangular part is equivalent to a transposed version of its upper triangular part, excluding the main diagonal).
Fig.~\ref{fig:matrix_structures}(c) shows an example sympd matrix.
For solving systems of linear equations,
the sympd property can be exploited by using Cholesky decomposition~\cite{Golub_2013} instead of LU decomposition
in order to reduce computational effort.

To determine whether a given matrix is sympd,
the traditional approach is to check whether all its eigenvalues are real and positive~\cite{Bhatia_2015}.
However, this raises two complications.
First, the eigen-decomposition of a matrix is computationally expensive,
which would defeat the aim of saving computational effort.
Second, since the matrix is stored in simple column-major format,
all matrix elements must be checked to ensure that symmetry is actually present.
This in turn raises a further complication, 
as a simple check for equality between elements
does not take into account minor differences that may be present
due to the accumulation of rounding errors
stemming from limitations in the precision of floating point representations~\cite{Goldberg_1991,Muller_2018}.

To address the above issues,
we have devised a fast two-pass algorithm which aims to
ensure the presence of several necessary conditions for a sympd matrix~\cite{Bhatia_2015,Golub_2013}:
{\bf (i)}~all diagonal elements are greater than zero,
{\bf (ii)}~the element with largest modulus is on the main diagonal,
{\bf (iii)}~the matrix is diagonally dominant,
and
{\bf (iv)}~the matrix is symmetric while tolerating minor variations in a robust manner.

We note that while the conditions checked by the algorithm are necessary,
they are not sufficient to absolutely guarantee the presence of a sympd matrix.
However, for practical purposes, 
in our experience (mainly in the machine learning area)
the conditions appear adequate for determining sympd presence.

The algorithm is shown in Fig.~\ref{fig:likely_sympd},
which returns either {\it true} or {\it false} to indicate that a matrix is likely to be sympd.
As soon as any condition is not satisfied, the algorithm aborts further processing and returns {\it false}.
Lines 7 to 8 check whether all diagonal elements are greater than zero.
Lines 12 to 14 check the presence of symmetry by determining whether the difference between an element and its corresponding transposed element is below a threshold;
the difference is compared against the threshold in both absolute and relative terms,
for robustness against precision variations between low-magnitude and high-magnitude floating point representations~\cite{Goldberg_1991,Muller_2018}.
We have empirically chosen the threshold as $100 \cdot \epsilon$,
where $\epsilon$ is the {\it machine epsilon}%
\footnote
  {
  Machine epsilon ($\epsilon$) is defined as the difference between 1 and the next representable floating point value~\cite{Goldberg_1991,Muller_2018}.
  For double-precision floating point numbers on {\tt x86-64} machines,
  {\small $\epsilon \approx 2.22045 \times 10^{-16}$}.
  }%
.
Line 15 checks whether the element with largest modulus is on the main diagonal,
while line 16 performs a rudimentary check for diagonal dominance.

If the algorithm determines that a sympd matrix is present,
an attempt to solve the system is made with a combination of the following LAPACK functions: {\tt xPOTRF}, {\tt xPOTRS}, and {\tt xPOCON}.
The {\tt xPOTRF} function computes the Cholesky decomposition,
{\tt xPOTRS} solves the system based on the Cholesky decomposition,
and finally {\tt xPOCON} computes the reciprocal condition number from the Cholesky decomposition.

If the {\tt xPOTRF} function fails (\ie the Cholesky decomposition was not found),
the failure is taken to indicate that the matrix is not actually sympd.
In that case, the system is solved by the generic solver described in Section~\ref{sec:gen_matrices}.

\begin{figure}[!b]
\vspace{-2ex}
\centering
\begin{adjustbox}{minipage=\columnwidth,scale={0.99}{1}}   
\begin{minipage}[t]{1\textwidth}
\hrule
\vspace{0.5ex}
\begin{tabbing}
AB\=A\=A\=A\=A\=A\=\kill
\texttt{~1}  \> {\bf proc likely\_sympd} \\
\texttt{~2}  \>\> {\bf input:} $A$ (square matrix in column-major format)\\
\texttt{~3}  \>\> {\bf input:} $N$ (number of rows in $A$) \\
\texttt{~4}  \>\> {\bf output:} boolean ({\it true} or {\it false}) \\
\texttt{~5}  \>\> $\text{\it tol} \leftarrow 100 \cdot \epsilon$ ~ (where $\epsilon$ is machine epsilon)\\
\texttt{~6}  \>\> $\text{\it max\_diag} \leftarrow 0$ \\
\texttt{~7}  \>\> {\bf forall} $j \in [0,N)$: \\
\texttt{~8}  \>\>\> {\bf if} $A_{j,j} \leq 0$ {\bf then return} {\it false} \\
\texttt{~9}  \>\>\> {\bf if} $A_{j,j} > \text{max\_diag}$ {\bf then} $\text{max\_diag} \leftarrow A_{j,j}$ \\
\texttt{10}  \>\> {\bf forall} $j \in [0,N-1)$: \\
\texttt{11}  \>\>\> {\bf forall} $i \in [j+1,N)$: \\
\texttt{12}  \>\>\>\> \text{\it delta} $\leftarrow | A_{i,j} - A_{j,i} |$ \\
\texttt{13}  \>\>\>\> {\bf if} \text{\it delta} $> \text{\it tol}$ {\bf and} \text{\it delta} $> \text{\it tol} \cdot \max(|A_{i,j}|, |A_{j,i}|)$ \\
\texttt{14}  \>\>\>\>\> {\bf then return} {\it false} \\
\texttt{15}  \>\>\>\> {\bf if} $|A_{i,j}| \geq \text{\it max\_diag}$ {\bf then return} {\it false} \\
\texttt{16}  \>\>\>\> {\bf if} $\left( |A_{i,j}| + |A_{j,i}| \right) \geq \left( A_{i,i} + A_{j,j} \right)$ {\bf then return} {\it false} \\
\texttt{17}  \>\> {\bf return} {\it true}
\end{tabbing}%
\hrule
\end{minipage}
\end{adjustbox}
\caption
  {
  An algorithm for detecting whether a given matrix $A$ is likely to be symmetric positive definite (sympd).
  The matrix is assumed to have column-major storage,
  with its elements accessed via $A_{i,j}$, where $i$ indicates the row and $j$ indicates the column.
  Indexing starts at zero, following C++ convention.
  }
\label{fig:likely_sympd}
\end{figure}

\subsection{General Matrices}
\label{sec:gen_matrices}

For general matrices, where no special structure has been detected, a standard LU-based solver is used.
Here matrix $\Mat{A}$ from Eqn.~(\ref{eqn:general_form}) is expressed as
$\Mat{A} = \Mat{L}\Mat{U}$,
where
$\Mat{L}$ is a unit lower triangular matrix,
and $\Mat{U}$ is an upper triangular matrix.
Rewriting Eqn.~(\ref{eqn:general_form}) yields:%
\begin{equation}
  \Mat{L}\left(\Mat{U}\Vec{x}\right) = \Vec{b}
  \label{eqn:lu_form}
\end{equation}

\noindent
Solving for $\Vec{x}$ is then accomplished in two steps.
In the first step, Eqn.~(\ref{eqn:lu_form}) is rewritten as $\Mat{L}\Mat{Y} = \Vec{b}$,
and a solution for $\Mat{Y}$ is found.
Since $\Mat{L}$ is lower triangular and $\Vec{b}$ is known,
$\Mat{Y}$ is easily found via forward-substitution (as shown in Section~\ref{sec:tri_matrices}).
In the second step, $\Mat{Y}$ is expressed as $\Mat{Y} = \Mat{U}\Vec{x}$.
Since $\Mat{U}$ is lower triangular and $\Mat{Y}$ is known,
$\Vec{x}$ is found via back-substitution,
thereby providing the desired solution to Eqn.~(\ref{eqn:general_form}).

Rather than simply using the {\tt xGESV} family of functions from LAPACK
that do not provide an estimate of the reciprocal condition number,
we use a combination of {\tt xGETRF}, {\tt xGETRS} and {\tt xGECON}.
The {\tt xGETRF} function computes the LU decomposition,
{\tt xGETRS} solves a system of linear equations based on the LU decomposition,
and finally {\tt xGECON} computes the reciprocal condition number from the LU decomposition.

\begin{table*}[t!]
\begin{minipage}{1\textwidth}
\centering
\begin{minipage}{0.48\textwidth}
\centering
\normalsize
\begin{tabular}{cccc}
\hline
{\bf matrix size} & {\bf standard}         & {\bf adaptive}         & {\bf reduction} \\ \hline
 100$\times$100   & $1.48 \times 10^{-4}$  & $4.76 \times 10^{-5}$  & 67.89\%         \\
 250$\times$250   & $1.00 \times 10^{-3}$  & $1.66 \times 10^{-4}$  & 83.45\%         \\
 500$\times$500   & $5.41 \times 10^{-3}$  & $5.94 \times 10^{-4}$  & 89.02\%         \\
1000$\times$1000  & $3.27 \times 10^{-2}$  & $2.89 \times 10^{-3}$  & 91.18\%         \\ \hline
\end{tabular}
\caption
  {
  Comparison of time taken (in seconds)
  to solve random banded systems using a standard dense solver (which ignores the banded property)
  against the adaptive solver (which takes into account the banded property).
  Average wall-clock time across 1000 runs is reported.
  }
\label{tab:results_5band}
\vspace{4.6ex}
\begin{tabular}{cccc}
\hline
{\bf matrix size} & {\bf standard}         & {\bf adaptive}         & {\bf reduction} \\ \hline
 100$\times$100   & $1.40 \times 10^{-4}$  & $3.82 \times 10^{-5}$  & 72.78\%         \\
 250$\times$250   & $1.05 \times 10^{-3}$  & $2.78 \times 10^{-4}$  & 73.59\%         \\
 500$\times$500   & $5.48 \times 10^{-3}$  & $1.12 \times 10^{-3}$  & 79.61\%         \\
1000$\times$1000  & $3.26 \times 10^{-2}$  & $5.19 \times 10^{-3}$  & 84.09\%         \\ \hline
\end{tabular}
\caption
  {
  Comparison of time taken (in seconds)
  to solve random triangular systems using a standard dense solver (which ignores the triangular property)
  against the adaptive solver (which takes into account the triangular property).
  Average wall-clock time across 1000 runs is reported.
  }
\label{tab:results_trimat}
\end{minipage}
\hfill
\begin{minipage}{0.48\textwidth}
\centering
\normalsize
\begin{tabular}{cccc}
\hline
{\bf matrix size} & {\bf standard}         & {\bf adaptive}         & {\bf reduction} \\ \hline
 100$\times$100   & $1.44 \times 10^{-4}$  & $1.19 \times 10^{-4}$  & 17.35\%         \\
 250$\times$250   & $1.03 \times 10^{-3}$  & $7.66 \times 10^{-4}$  & 25.62\%         \\
 500$\times$500   & $5.63 \times 10^{-3}$  & $4.27 \times 10^{-3}$  & 24.02\%         \\
1000$\times$1000  & $3.31 \times 10^{-2}$  & $2.40 \times 10^{-2}$  & 27.54\%         \\ \hline
\end{tabular}
\caption
  {
  Comparison of time taken (in seconds)
  to solve random symmetric positive definite (sympd) systems using a standard dense solver (which ignores the sympd property)
  against the adaptive solver (which takes into account the sympd property).
  Average wall-clock time across 1000 runs is reported.
  }
\label{tab:results_sympd}
\vspace{2ex}
\begin{tabular}{cccc}
\hline
{\bf matrix size} & {\bf standard}             & {\bf adaptive}             & {\bf overhead} \\ \hline
 100$\times$100   & $ 1.486 \times 10^{-4}$  & $ 1.511 \times 10^{-4}$  & 1.627\%         \\
 250$\times$250   & $ 1.220 \times 10^{-3}$  & $ 1.223 \times 10^{-3}$  & 0.243\%         \\
 500$\times$500   & $ 6.056 \times 10^{-3}$  & $ 6.063 \times 10^{-3}$  & 0.114\%         \\
1000$\times$1000  & $ 3.598 \times 10^{-2}$  & $ 3.605 \times 10^{-2}$  & 0.187\%         \\ \hline
\end{tabular}
\caption
  {
  Comparison of time taken (in seconds)
  to solve random dense systems (without any special structure) using a standard dense solver
  against the adaptive solver (which attempts to detect special structures).
  Average wall-clock time across 1000 runs is reported.
  }
\label{tab:results_dense}
\end{minipage}
\end{minipage}
\vspace{2ex}
\end{table*}

\subsection{Fallback for Poorly Conditioned Systems}
\label{sec:svd_solver}

If the above solvers fail or if the estimated reciprocal condition number indicates a poorly conditioned system,
an approximate solution is attempted using a solver based on SVD.
A poorly conditioned system is detected when the reciprocal condition number is below a threshold;
following LAPACK's example, we have set this threshold to {\small $0.5\epsilon$}, 
where $\epsilon$ is the machine epsilon as defined in Section~\ref{sec:sympd_matrices}.
The use of the SVD-based fallback solver can be disabled through an optional argument to the {\tt solve()} function.

The SVD-based solver uses the {\tt xGELSD} set of functions,
which find a minimum-norm solution to a linear least squares problem~\cite{Boyd_2018},
\ie $\Vec{x}$ is found via minimisation of {\small $\left|\left| \Vec{b} - \Mat{A} \Vec{x} \right|\right|_{2}$}.
In brief, the solution to the system in Eqn.~(\ref{eqn:general_form}) is reformulated as
\mbox{\small $\Vec{x} = (\Mat{A}^{\top} \Mat{A})^{-1} \Mat{A}^{\top} \Vec{b}$},
where
\mbox{\small $(\Mat{A}^{\top} \Mat{A})^{-1} \Mat{A}^{\top}$}
is known as the Moore-Penrose pseudo-inverse,
which can be obtained via the SVD of $\Mat{A}$~\cite{Golub_2013}.
This reformulation is equivalent to the least squares solution.

\section{Empirical Evaluation}
\label{sec:experiments}
\vspace{0.5ex}

In this section we demonstrate the speedups resulting from 
automatically detecting the matrix properties described in Section~\ref{sec:mapping}
and selecting the most appropriate set of LAPACK functions for solving systems of linear equations.
The evaluations were done on a machine with an Intel Core i5-5200U CPU running at 2.2~GHz.
Compilation was done with the GCC 10.1 C++ compiler with the following configuration options: {\tt -O3 -march=native}.
We used the open-source OpenBLAS~0.3.9~\cite{OpenBLAS} package which provides optimised implementations of LAPACK functions.

We performed evaluations on the following set of sizes for matrix $\Mat{A}$,
ranging from small to large:
{\small \{100$\times$100,  250$\times$250, 500$\times$500, 1000$\times$1000\}}.
Three matrix types were used:
{\bf (i)}~banded (with 5 diagonals),
{\bf (ii)}~lower triangular,
{\bf (iii)}~symmetric positive definite (sympd).
For each matrix size and type, 1000 unique random systems were solved,
with each random system solved using the standard LU-based solver (as per Section~\ref{sec:gen_matrices})
and the adaptive solver.
For each solver, the average wall-clock time (in seconds) across the 1000 runs is reported.
The results are presented in Tables~\ref{tab:results_5band}, \ref{tab:results_trimat} and~\ref{tab:results_sympd}.

The results indicate that for all matrix types,
the adaptive solver reduces the wall-clock time.
On average, the reduction is most notable for banded systems,
closely followed by triangular systems.
While the reductions for sympd systems are less pronounced,
they still show useful speedups.
In general, the larger the matrix size, the larger the degree in reduction of wall-clock time.

To gauge the overhead of all the detection algorithms,
we compare the time taken to solve random dense systems (without any special structure)
using the standard LU-based solver against the adaptive solver.
The results shown in Table~\ref{tab:results_dense} indicate that the overhead is negligible.
This is due to each detection algorithm stopping as soon as it determines 
that a special structure is not present.

\vspace{1ex}
\section{Runtime Considerations}
\label{sec:runtime}

In the previous section we showed the proposed algorithm to be empirically
efficient on randomly generated linear systems.
Indeed, this is partly due to the fact that the checks that we have proposed are
asymptotically more efficient than the factorisations we employ.
To see this, observe that the banded structure check described in
Section~\ref{sec:banded_matrices} can be performed in a single pass over the matrix.
For a square matrix of size {\small $n \times n$}, this is {\small $O(n^2)$} time.
Similarly, the triangular structure check described in Section~\ref{sec:tri_matrices}
can also be performed in a single pass, yielding {\small $O(n^2)$} time.
Finally, the {\bf likely\_sympd} algorithm in Fig.~\ref{fig:likely_sympd},
as written, can be performed as an {\small $O(n)$} diagonal pass on {\small $\Mat{A}$}
followed by a single {\small $O(n^2)$} pass.
Overall, this yields a total quadratic runtime for all of the proposed checks.

In the situation where the given matrix is triangular,
we can perform forward- or back-substitution in {\small $O(n^2)$} time,
yielding an asymptotic improvement over the standard LU-based solver,
which takes $O(n^3)$ time~\cite{Golub_2013}.
Cholesky decomposition and singular value decomposition both also take
$O(n^3)$ time to compute~\cite{Golub_2013}.
However, in practice the Cholesky decomposition tends to be faster
(as only one triangular part of the symmetric matrix needs to be processed),
explaining the rest of the empirical advantage that was observed in our experiments.

\section{Conclusion}
\label{sec:conclusion}
\vspace{1ex}

We have described an adaptive solver for systems of linear equations
that is able to automatically detect several common properties of a given system
(banded, triangular, and symmetric positive definite),
followed by solving the system via mapping to a set of suitable LAPACK functions best matched to each property.
Furthermore, the solver can detect poorly conditioned systems and automatically obtain an approximate solution
via singular value decomposition as a fallback.
The automatic handling facilitates simplified user code that focuses on high-level algorithm logic,
freeing the user from worrying about various storage formats
and using cumbersome manual calls to LAPACK functions in
order to obtain good computational efficiency.

The solver is present inside recent versions of the high-level Armadillo C++ library for linear algebra~\cite{Armadillo_JOSS_2016,armadillo_spmat_2019},
which allows matrix operations to be expressed in an easy-to-read manner similar to Matlab/Octave.
The solver is comprised of about 3000 lines of code, not counting LAPACK code;
it is able to handle matrices with single- and double-precision floating point elements, in both real and complex forms.
The source code for the adaptive solver is provided under the permissive Apache~2.0 license~\cite{Laurent_2008},
allowing unencumbered use in commercial products;
it can be obtained from \href{http://arma.sourceforge.net}{\small\tt http://arma.sourceforge.net}.

The adaptive solver has been successfully used to increase efficiency in open-source toolkits
such as the {\it mlpack} library for machine learning~\cite{mlpack2018},
and the {\it ensmallen} library for numerical optimisation~\cite{ensmallen2018}.
Areas for further improvements include specialised handling
of plain symmetric matrices (symmetric but not positive definite),
and tri-diagonal matrices which are a common subtype of banded matrices~\cite{Cheney_2012}.

\vspace{1ex}
\section*{Acknowledgements}
\vspace{1ex}

We would like to thank our colleagues at Data61/CSIRO 
(Frank De Hoog, Mark Staples)
and Griffith University
(M.A.~Hakim Newton, Majid Namazi)
for discussions leading to the improvements of this paper.

\newpage
\bibliographystyle{ieee}
\bibliography{refs}
\clearpage

\end{document}